\documentclass[a4paper,11pt]{article}
\usepackage[dvipsnames]{xcolor}

\usepackage{pos}
\usepackage{multirow}
\usepackage{amsmath,amsthm,amsfonts}
\usepackage{caption,subcaption}
\usepackage{tikz,tikz-3dplot}
\usepackage[compat=1.1.0]{tikz-feynman}
\usetikzlibrary{calc,arrows.meta,shapes.geometric}

\def\I{\mathcal{I}}

\def\x{\boldsymbol{x}}
\def\v{\boldsymbol{v}}

\def\s{\boldsymbol{s}}

\def\eps{\epsilon}

\title{Revealing Hidden Regions in Wide-Angle and Forward Scattering}

\author[a]{Einan Gardi} 
\author[a]{Franz Herzog}
\author*[b]{Stephen Jones}
\author[c]{Yao Ma}

\affiliation[a]{Higgs Centre for Theoretical Physics, School of Physics and Astronomy,\\The University of Edinburgh, Edinburgh EH9 3FD, Scotland, UK}
\affiliation[b]{Institute for Particle Physics Phenomenology, Durham University, Durham DH1 3LE, UK}
\affiliation[c]{Institute for Theoretical Physics, ETH Zürich, 8093 Zürich, Switzerland}

\emailAdd{einan.gardi@ed.ac.uk}
\emailAdd{fherzog@ed.ac.uk}
\emailAdd{stephen.jones@durham.ac.uk}
\emailAdd{yaomay@phys.ethz.ch}

\abstract{
We discuss a class of Feynman Integrals containing hidden regions that are not straightforwardly identified using the geometric, or Newton polytope, approach to the method of regions. 
Using Landau singularity analysis and existing analytic results, we study the appearance of such regions in wide-angle and forward scattering and discuss how they can be exposed in both the momentum and parametric representations. 
We demonstrate that in the strict on-shell limit such integrals contain Landau singularities that prevent their direct numerical evaluation in parameter space and describe how they can be re-parameterised and dissected to circumvent this problem.
}

\FullConference{Loops and Legs in Quantum Field Theory (LL2024)\\
 14-19, April, 2024\\
Wittenberg, Germany\\}

\begin{document}
\maketitle

\section{Introduction}

Evaluating Feynman integrals is an essential and often challenging step in computing higher-order perturbative corrections in Quantum Field Theories.
In part due to the pioneering work of Landau, Bjorken and Nakanishi~\cite{Lnd59,Bjorken:1959fd,Nakanishi,EdenLdshfOlvPkhn02book}, the study of the analytic structure of Feynman integrals as dictated by their singularities, has played a central role in devising methods for their evaluation both analytically and numerically.
In recent years, the study of Landau singularities and the Landau equations has experienced a revival of interest, leading to new and more efficient ways to compute and characterise solutions~\cite{Brown:2009ta,Panzer:2014caa,GrdHzgJnsMaSchlk22,FvlMzrTln23pld,FvlMzrTln23prl,Dlapa:2023cvx,Helmer:2024wax}.

One promising direction of exploration is the geometric description of Feynman integrals in parameter space, i.e. in Feynman or Lee-Pomeransky representation~\cite{LeePmrsk13}.
In this representation, each monomial of the Symanzik polynomials, $\mathcal{U}$ and $\mathcal{F}$,
characterising a Feynman graph with $N$ edges is mapped to an $N$-dimensional vector whose entries are the exponents of the parameters in that monomial.
Regarding each of these vectors as a point in an $N$-dimensional space, a Newton polytope can be obtained by taking the convex hull of all these points. 
The normal vectors to the faces of the polytope are then precisely the weight vectors characterising the endpoint divergences of the integral, which appear when some set of parameters vanish or become large. 
This geometric description facilitates the formulation of efficient algorithms to determine Landau singularities~\cite{Brown:2009ta,Panzer:2014caa,FvlMzrTln23pld,FvlMzrTln23prl,Dlapa:2023cvx,Helmer:2024wax}, the study of the infrared singularity structure of Feynman integrals in dimensional regularisation~\cite{AkHmHlmMzr22}, the numerical computation Feynman integrals using sector decomposition~\cite{SmnTtyk09FIESTA,SmnSmnTtyk11FIESTA2,Smn14FIESTA3,Smn16FIESTA4,Smn22FIESTA5,KnkUeda10,Carter:2010hi,Borowka:2015mxa,Heinrich:2023til,pySecDec17,Brsk20,BrskMchTld23,LiuMa23AMFlow,Hdg21DiffExp,AmdlBcnDvtRanaVcn23SeaSyde}, or tropical Monte Carlo integration~\cite{Brsk20,BrskMchTld23}, and the asymptotic expansion of Feynman integrals by the method of regions (MoR) directly in parameter space as implemented in various computer packages~\cite{JtzSmnSmn12,AnthnrySkrRmn19,pySecDec17,HrchJnsSlk22}.

However, not all singularities, or solutions of the Landau equations, appear as endpoint divergences in parameter space.
Indeed, the necessary conditions for a singularity to occur are summarised by the Landau equations, which can be written in parameter space as~\cite{Lnd59,EdenLdshfOlvPkhn02book},
\begin{align}
\mathcal{F}&=0 \\
\alpha_k \, \frac{\partial\mathcal{F}}{\partial \alpha_k} &=0\quad \text{for each }k\in\{1,\dots,N\},
\label{eq:Landau_equation_parameter_representation}
\end{align}
where $\mathcal{F}$ is the second Symanzik polynomial.
Examining these equations, we can see that solutions, known as pinch solutions, can occur when both ${\cal F}$ and $\partial {\cal F}/\partial \alpha_i$ vanish away from the integration boundary. 
Although this situation has been considered since the early days, its consequences for evaluating Feynman integrals using sector decomposition and applying the MoR in parameter space have not yet been fully explored.

Parameter space pinch solutions of the Landau equations occur when monomials of opposite sign in the $\mathcal{F}$ polynomial cancel each other in such a way that $\mathcal{F}$ and its derivatives $\partial {\cal F}/\partial \alpha_i$ vanish.
The Newton polytope is completely insensitive to the coefficients of the monomials (including their sign), so such singularities can not be identified from the polytope alone.
Solutions can arise in special kinematic limits, such as intermediate particle thresholds or forward limits, giving rise to hidden regions, e.g. potential and Glauber regions, respectively, when using the MoR to expand integrals about such limits.
Here we discuss cases where such solutions are present even with general kinematics (i.e. not in threshold or forward limit).

In some cases, hidden regions can be found algorithmically by selecting a variable appearing in a positive monomial, $\alpha_i$, and a variable appearing in a negative monomial, $\alpha_j$, then making iterated linear changes of variables chosen to eliminate negative monomials in $\mathcal{F}$.
This procedure was proposed and implemented in the Asy2 package in Ref.~\cite{JtzSmnSmn12}.
It has also been applied instead to the Gr\"{o}bner basis of the $\mathcal{F}$ polynomial and its first order derivatives in Ref.~\cite{AnthnrySkrRmn19}, which introduced the ASPIRE package.
We explore cases where iterated linear changes of variables do not eliminate all negative monomials in $\mathcal{F}$ and fail to expose hidden regions which appear even at leading power in the expansion.

These proceedings provide a short overview of the main results of the work presented in Ref.~\cite{Gardi:2024axt}, where the structure of pinch solutions of the Landau equations for general kinematics is analysed in the case of $2 \rightarrow 2$ massless scattering.
It is observed that such solutions prevent the direct numerical evaluation of certain integrals, starting at three loops, and that they are associated with hidden regions that can not be resolved by iterated linear changes of variables.
In Section~\ref{sec:identifying} we briefly describe how integrals that may contain cancellation structures can be found systematically.
In Section~\ref{sec:evaluating} we discuss how pinch singularities can prevent the numerical evaluation of Feynman integrals and give a non-trivial example of how re-parameterising and dissecting such integrals can circumvent this problem.
We demonstrate, in Section~\ref{sec:hidden}, how hidden regions can be found both in momentum and parameter space.
Finally, we present our conclusions in Section~\ref{sec:conclusions}.

\section{Identifying Integrals with Pinch Singularities}
\label{sec:identifying}

For a parameter space pinch solution to be present in a Feynman integral, we require that both $\mathcal{F}$ and $\partial \mathcal{F}/\partial \alpha_i$ vanish for some $\alpha_i \neq 0$.
To identify integrals for which this is possible, we begin by defining the polynomials $\mathcal{F}_+$ and $\mathcal{F}_-$ as the positive and negative monomials of $\mathcal{F}$ at a given kinematic point, respectively, such that $\mathcal{F}=\mathcal{F}_+ + \mathcal{F}_-$.
At a pinch singularity, some terms of $\mathcal{F}_+$ cancel against terms of $\mathcal{F}_-$, both in $\mathcal{F}$ and (some of) its first-order partial derivatives.
When searching for this cancellation with general kinematics (i.e. independently of the specific values chosen for the independent set of Mandelstam invariants), as we do here, we can search for the pinch solutions invariant-by-invariant, considering only terms in $\mathcal{F}$ proportional to a single Mandelstam invariant.
Let us denote the terms in $\mathcal{F}$ that are proportional to the invariant $s_{ij}$ by $\mathcal{F}^{(s_{ij})}$.
We can then formulae a search algorithm as follows:
\begin{itemize}
    \item [] 1. Compute $\mathcal{F}^{(s_{ij})}_+$ and $\mathcal{F}^{(s_{ij})}_-$. If either vanishes, exit the algorithm outputting that there are no pinch solution. Otherwise, go to Step 2.
    \item [] 2. Compute $\partial\mathcal{F}^{(s_{ij})}_+/\partial \alpha_i$ and $\partial\mathcal{F}^{(s_{ij})}_-/\partial \alpha_i$ for all the $\alpha_i$ that $\mathcal{F}^{(s_{ij})}$ depends on. If none of these derivatives vanish, exit the algorithm outputting that there are pinch singularities (which may or may not be within the integration domain). Otherwise, go to Step 3.
    \item [] 3. Identify the $i$ for which~$\partial\mathcal{F}^{(s_{ij})}_+/\partial \alpha_i = 0$ or $\partial\mathcal{F}^{(s_{ij})}_-/\partial \alpha_i = 0$, replace $\mathcal{F}^{(s_{ij})}$ by $\mathcal{F}^{(s_{ij})}|_{\alpha_i=0}$, and return to Step 1.
\end{itemize}
For further discussion and justification of this algorithm, we refer the reader to Ref.~\cite{Gardi:2024axt}.

\begin{figure}[t]
\centering
\begin{subfigure}[b]{0.25\textwidth}
\centering
\resizebox{\textwidth}{!}{
\begin{tikzpicture}[line width = 0.6, font=\large, mydot/.style={circle, fill, inner sep=.7pt}]
% \draw [help lines] (0,0) grid (10,10);

\draw (0.5,1) edge [ultra thick, color=Green] (2,2) node [] {};
\draw (9.5,1) edge [ultra thick, color=teal] (8,2) node [] {};
\draw (0.5,9) edge [ultra thick, color=LimeGreen] (2,8) node [] {};
\draw (9.5,9) edge [ultra thick, color=olive] (8,8) node [] {};
\draw (5,7.5) edge [ultra thick, bend right = 10] (2,2) node [] {};
\draw (5,7.5) edge [ultra thick, bend left = 10] (8,2) node [] {};
\draw (5,7.5) edge [ultra thick, bend right = 10] (2,8) node [] {};
\draw (5,7.5) edge [ultra thick, bend left = 10] (8,8) node [] {};
\draw (5,2.5) edge [ultra thick, draw=white, double=white, double distance=3pt, bend left = 10] (2,2) node [] {};\draw (5,2.5) edge [ultra thick, bend left = 10] (2,2) node [] {};
\draw (5,2.5) edge [ultra thick, draw=white, double=white, double distance=3pt, bend right = 10] (8,2) node [] {};\draw (5,2.5) edge [ultra thick, bend right = 10] (8,2) node [] {};
\draw (5,2.5) edge [ultra thick, draw=white, double=white, double distance=3pt, bend left = 10] (2,8) node [] {};\draw (5,2.5) edge [ultra thick, bend left = 10] (2,8) node [] {};
\draw (5,2.5) edge [ultra thick, draw=white, double=white, double distance=3pt, bend right = 10] (8,8) node [] {};\draw (5,2.5) edge [ultra thick, bend right = 10] (8,8) node [] {};

\node () at (0.5,0.5) {\huge $p_1$};
\node () at (9.5,0.5) {\huge $p_3$};
\node () at (0.5,9.5) {\huge $p_2$};
\node () at (9.5,9.5) {\huge $p_4$};

\node () at (3.5,1.5) {\huge $1$};
\node () at (2,3.5) {\huge $0$};
\node () at (8,3.5) {\huge $4$};
\node () at (6.5,1.5) {\huge $5$};
\node () at (8,6.5) {\huge $7$};
\node () at (6.5,8.5) {\huge $6$};
\node () at (3.5,8.5) {\huge $2$};
\node () at (2,6.5) {\huge $3$};

\draw[fill, thick] (2,2) circle (3pt);
\draw[fill, thick] (8,2) circle (3pt);
\draw[fill, thick] (2,8) circle (3pt);
\draw[fill, thick] (8,8) circle (3pt);
\draw[fill, thick] (5,7.5) circle (3pt);
\draw[fill, thick] (5,2.5) circle (3pt);
\end{tikzpicture}
}
\vspace{-3em}\caption{$G_{\bullet \bullet}$}
\label{sfig:g_dot_dot}
\end{subfigure}
\\
\begin{subfigure}[b]{0.25\textwidth}
\centering
\resizebox{\textwidth}{!}{
\begin{tikzpicture}[line width = 0.6, font=\large, mydot/.style={circle, fill, inner sep=.7pt}]
% \draw [help lines] (0,0) grid (10,10);

\draw (0.5,1) edge [ultra thick] (2,2) node [] {};
\draw (9.5,1) edge [ultra thick] (8,2) node [] {};
\draw (0.5,9) edge [ultra thick] (2,8) node [] {};
\draw (9.5,9) edge [ultra thick] (8,8) node [] {};
\draw (4,2.5) edge [ultra thick] (6,2.5) node [] {};
\draw (5,7.5) edge [ultra thick, bend right = 10] (2,2) node [] {};
\draw (5,7.5) edge [ultra thick, bend left = 10] (8,2) node [] {};
\draw (5,7.5) edge [ultra thick, bend right = 10] (2,8) node [] {};
\draw (5,7.5) edge [ultra thick, bend left = 10] (8,8) node [] {};
\draw (4,2.5) edge [ultra thick, draw=white, double=white, double distance=3pt, bend left = 10] (2,2) node [] {};\draw (4,2.5) edge [ultra thick, bend left = 10] (2,2) node [] {};
\draw (6,2.5) edge [ultra thick, draw=white, double=white, double distance=3pt, bend right = 10] (8,2) node [] {};\draw (6,2.5) edge [ultra thick, bend right = 10] (8,2) node [] {};
\draw (4,2.5) edge [ultra thick, draw=white, double=white, double distance=3pt, bend left = 10] (2,8) node [] {};\draw (4,2.5) edge [ultra thick, bend left = 10] (2,8) node [] {};
\draw (6,2.5) edge [ultra thick, draw=white, double=white, double distance=3pt, bend right = 10] (8,8) node [] {};\draw (6,2.5) edge [ultra thick, bend right = 10] (8,8) node [] {};

\node () at (0.5,0.5) {\huge $p_1$};
\node () at (9.5,0.5) {\huge $p_3$};
\node () at (0.5,9.5) {\huge $p_2$};
\node () at (9.5,9.5) {\huge $p_4$};

\draw[fill, thick] (2,2) circle (3pt);
\draw[fill, thick] (8,2) circle (3pt);
\draw[fill, thick] (2,8) circle (3pt);
\draw[fill, thick] (8,8) circle (3pt);
\draw[fill, thick] (5,7.5) circle (3pt);
\draw[fill, thick] (4,2.5) circle (3pt);
\draw[fill, thick] (6,2.5) circle (3pt);

\end{tikzpicture}
}
\vspace{-3em}\caption{$G_{\bullet s}$}
\label{sfig:g_dot_s}
\end{subfigure}
\qquad
\begin{subfigure}[b]{0.25\textwidth}
\centering
\resizebox{\textwidth}{!}{
\begin{tikzpicture}[line width = 0.6, font=\large, mydot/.style={circle, fill, inner sep=.7pt}]
% \draw [help lines] (0,0) grid (10,10);

\draw (0.5,1) edge [ultra thick] (2,2) node [] {};
\draw (9.5,1) edge [ultra thick] (8,2) node [] {};
\draw (0.5,9) edge [ultra thick] (2,8) node [] {};
\draw (9.5,9) edge [ultra thick] (8,8) node [] {};
\draw (4,4) edge [ultra thick] (4,6) node [] {};
\draw (6,4) edge [ultra thick] (6,6) node [] {};
\draw (6,4) edge [ultra thick, bend left = 10] (2,2) node [] {};
\draw (6,4) edge [ultra thick, bend left = 10] (8,2) node [] {};
\draw (6,6) edge [ultra thick, bend right = 10] (2,8) node [] {};
\draw (6,6) edge [ultra thick, bend right = 10] (8,8) node [] {};
\draw (4,4) edge [ultra thick, draw=white, double=white, double distance=3pt, bend right = 10] (2,2) node [] {};\draw (4,4) edge [ultra thick, bend right = 10] (2,2) node [] {};
\draw (4,4) edge [ultra thick, draw=white, double=white, double distance=3pt, bend right = 10] (8,2) node [] {};\draw (4,4) edge [ultra thick, bend right = 10] (8,2) node [] {};
\draw (4,6) edge [ultra thick, draw=white, double=white, double distance=3pt, bend left = 10] (2,8) node [] {};\draw (4,6) edge [ultra thick, bend left = 10] (2,8) node [] {};
\draw (4,6) edge [ultra thick, draw=white, double=white, double distance=3pt, bend left = 10] (8,8) node [] {};\draw (4,6) edge [ultra thick, bend left = 10] (8,8) node [] {};

\node () at (0.5,0.5) {\huge $p_1$};
\node () at (9.5,0.5) {\huge $p_3$};
\node () at (0.5,9.5) {\huge $p_2$};
\node () at (9.5,9.5) {\huge $p_4$};

\draw[fill, thick] (2,2) circle (3pt);
\draw[fill, thick] (8,2) circle (3pt);
\draw[fill, thick] (2,8) circle (3pt);
\draw[fill, thick] (8,8) circle (3pt);
\draw[fill, thick] (6,6) circle (3pt);
\draw[fill, thick] (6,4) circle (3pt);
\draw[fill, thick] (4,6) circle (3pt);
\draw[fill, thick] (4,4) circle (3pt);

\end{tikzpicture}
}
\vspace{-3em}\caption{$G_{t t}$}
\label{sfig:g_t_t}
\end{subfigure}
\qquad
\begin{subfigure}[b]{0.25\textwidth}
\centering
\resizebox{\textwidth}{!}{
\begin{tikzpicture}[line width = 0.6, font=\large, mydot/.style={circle, fill, inner sep=.7pt}]
% \draw [help lines] (0,0) grid (10,10);

\draw (0.5,1) edge [ultra thick] (2,2) node [] {};
\draw (9.5,1) edge [ultra thick] (8,2) node [] {};
\draw (0.5,9) edge [ultra thick] (2,8) node [] {};
\draw (9.5,9) edge [ultra thick] (8,8) node [] {};
\draw (2.5,5) edge [ultra thick] (7.5,5) node [] {};
\draw (5,7.5) edge [ultra thick, draw=white, double=white, double distance=3pt] (5,2.5) node [] {};\draw (5,7.5) edge [ultra thick] (5,2.5) node [] {};
\draw (2.5,5) edge [ultra thick, bend right = 10] (2,2) node [] {};
\draw (5,7.5) edge [ultra thick, bend left = 10] (8,8) node [] {};
\draw (2.5,5) edge [ultra thick, bend left = 10] (2,8) node [] {};
\draw (7.5,5) edge [ultra thick, bend right = 10] (8,8) node [] {};
\draw (2,2) edge [ultra thick, draw=white, double=white, double distance=3pt, bend right = 10] (5,2.5) node [] {};\draw (2,2) edge [ultra thick, bend right = 10] (5,2.5) node [] {};
\draw (5,2.5) edge [ultra thick, draw=white, double=white, double distance=3pt, bend right = 10] (8,2) node [] {};\draw (5,2.5) edge [ultra thick, bend right = 10] (8,2) node [] {};
\draw (5,7.5) edge [ultra thick, draw=white, double=white, double distance=3pt, bend right = 10] (2,8) node [] {};\draw (5,7.5) edge [ultra thick, bend right = 10] (2,8) node [] {};
\draw (8,2) edge [ultra thick, draw=white, double=white, double distance=3pt, bend right = 10] (7.5,5) node [] {};\draw (8,2) edge [ultra thick, bend right = 10] (7.5,5) node [] {};

\node () at (0.5,0.5) {\huge $p_1$};
\node () at (9.5,0.5) {\huge $p_3$};
\node () at (0.5,9.5) {\huge $p_2$};
\node () at (9.5,9.5) {\huge $p_4$};

\draw[fill, thick] (2,2) circle (3pt);
\draw[fill, thick] (8,2) circle (3pt);
\draw[fill, thick] (2,8) circle (3pt);
\draw[fill, thick] (8,8) circle (3pt);
\draw[fill, thick] (2.5,5) circle (3pt);
\draw[fill, thick] (7.5,5) circle (3pt);
\draw[fill, thick] (5,7.5) circle (3pt);
\draw[fill, thick] (5,2.5) circle (3pt);

\end{tikzpicture}
}
\vspace{-3em}\caption{$G_{s t}$}
\label{sfig:g_s_t}
\end{subfigure}
\caption{Massless four-point three-loop graphs with a possible pinch Landau singularity in parameter space. Diagrams related by crossing are also identified as possibly containing pinch Landau singularities in parameter space.}
\label{fig:graphs_hidden}
\end{figure}
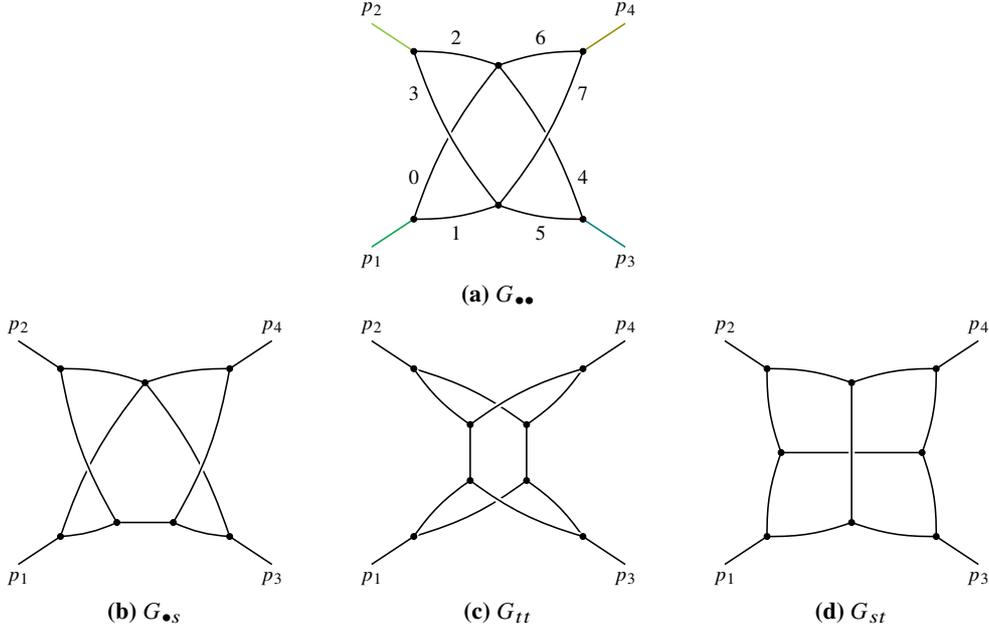

Running the algorithm on all one- and two-loop massless $2 \rightarrow 2$ graphs containing only three- and four-point vertices we find no integrals with pinch solutions.
Starting from three loops we identify 10 graphs for which a pinch solution is possible, a selection is shown in Figure~\ref{fig:graphs_hidden}.
We note that, remarkably, these graphs have been studied in the early literature in the context of Regge cuts~\cite{Mdst63}, the high-energy limit~\cite{Hld64} and elastic scattering~\cite{Ldsf74,BottsStm89}.
In Ref.~\cite{Gardi:2024axt}, we also report the four-loop graphs potentially containing pinch Landau singularities and find that for wide-angle scattering they all contain the aforementioned three-loop diagrams as subgraphs.

\section{Evaluating Integrals with a Pinched Contour}
\label{sec:evaluating}

In the Minkowski regime, Feynman integrals can have poles for real non-zero values of the parameters $\alpha_i$.
Such poles appear due to the vanishing of the $\mathcal{F}$ polynomial within the domain of integration of the Feynman parameters.
In typical cases, such singularities can be avoided by analytic continuation, or, in parameter space, by deforming the contour of integration of the $\alpha_k$-parameters into the complex plane by introducing a small imaginary part $\alpha_k\to \alpha_k - i\varepsilon_k (\boldsymbol{\alpha})$
According to the Feynman prescription, the deformation should be chosen such that $\mathcal{F}$ develops a negative imaginary part when the real part vanishes.
Under such a change of variables, the $\mathcal{F}$ polynomial transforms as
\begin{equation}
\mathcal{F}(\boldsymbol{\alpha};\s)\to \mathcal{F}(\boldsymbol{\alpha};\s) 
-i \sum_k \varepsilon_k (\boldsymbol{\alpha}) \frac{\partial \mathcal{F}(\boldsymbol{\alpha};\s)}{\partial \alpha_k}+\mathcal{O}(\varepsilon^2)\,.
\label{eq:contour_deformation}
\end{equation}
In the case of a pinch singularity, both $\mathcal{F}$ and $\partial \mathcal{F}/\partial \alpha_k$ vanish at the same point, pinching the contour and forcing it to vanish exactly where the deformation would be needed.
This prevents the direct numerical evaluation of the integral in parameter space.

An example Feynman integral containing a pinch singularity is given by $G_{\bullet\bullet}$, shown in Figure~\ref{sfig:g_dot_dot}. 
In momentum space, the integral can be written as,
\begin{align}
\label{eq:igbulletbullet}
\mathcal{I}_{G_{\bullet \bullet}}(\s) =& \int \left( \prod_{i=1}^3 \frac{d^Dk_i}{i\pi^{D/2}} \right) \left( \prod_{i=1}^3 \frac{1}{k_i^2 (k_i-p_i)^2} \right)
\frac{1}{(k_1+k_2-k_3)^2 (p_4-k_1-k_2+k_3)^2}, \nonumber\\
=& \Gamma(2+3\epsilon) \int d\alpha_0\dots d\alpha_7 \,\,
\delta(1-\alpha_0\dots -\alpha_7) \,
\Big(
\mathcal{U}(\boldsymbol{\alpha})\Big)^{4\eps}\,
\Big(\mathcal{F}(\boldsymbol{\alpha};\s)\Big)^{-2-3\eps},
\end{align}
the Symanzik polynomials read,
\begin{align}
\label{eq:UF_Gbb}
\begin{split}
    &\mathcal{U}(\boldsymbol{\alpha}) = (\alpha_0+\alpha_1)(\alpha_2+\alpha_3)(\alpha_4+\alpha_5) + (\alpha_0+\alpha_1)(\alpha_2+\alpha_3)(\alpha_6+\alpha_7)\\
    &\qquad\qquad + (\alpha_0+\alpha_1)(\alpha_4+\alpha_5)(\alpha_6+\alpha_7) + (\alpha_2+\alpha_3)(\alpha_4+\alpha_5)(\alpha_6+\alpha_7), \\
    &\mathcal{F}(\boldsymbol{\alpha};\s) =  (-s_{12}) (\alpha_1\alpha_4 - \alpha_0\alpha_5)(\alpha_3\alpha_6 - \alpha_2\alpha_7) + (-s_{13})(\alpha_1\alpha_2 - \alpha_0\alpha_3)(\alpha_5\alpha_6 - \alpha_4\alpha_7)\,,
\end{split}
\end{align}
where $s_{12}\equiv (p_1+p_2)^2$, $s_{13}\equiv  (p_1-p_3)^2$, and $s_{14} \equiv (p_1-p_4)^2$.
The momentum conservation relation $s_{12} + s_{13} + s_{23} = 0$ has been used to eliminate $s_{23}$, leading to monomials of different sign multiplying the remaining invariants $s_{12}$ and $s_{13}$.
The structure of the $\mathcal{F}$ polynomial is such that there is no analytic continuation prescription of the form $s_{ij} \rightarrow s_{ij} \pm i\varepsilon$ which ensures that $\mathcal{F}$ obtains a negative imaginary part for all values of the Feynman parameters $\alpha_i \geqslant 0$.

Possible pinch singularities in $\mathcal{F}$ of eq.~(\ref{eq:UF_Gbb}), are associated with the vanishing of subsets of the following polynomials,
\begin{align}
\label{v_iDef}
\begin{split}
    v_1=\alpha_1\alpha_4 - \alpha_0\alpha_5\,,\qquad v_2=\alpha_3\alpha_6 - \alpha_2\alpha_7\,,\\
    v_3=\alpha_1\alpha_2 - \alpha_0\alpha_3\,,\qquad v_4=\alpha_5\alpha_6 - \alpha_4\alpha_7\,.
\end{split}
\end{align}
The Landau equations can be satisfied with all $\alpha_i > 0$ for generic $s_{12}, s_{13}$ if and only if $v_1 = v_2 = v_3 = v_4 =0$, i.e. on the hypersurface defined by,
\begin{align}
\label{PinchSolution}
& \alpha_2 = \frac{\alpha_0\alpha_3}{\alpha_1}, &
& \alpha_4 = \frac{\alpha_0\alpha_5}{\alpha_1}, &
& \alpha_6 = \frac{\alpha_0\alpha_7}{\alpha_1}. &
\end{align}
These relations define a pinch surface.

We find that it is not possible to eliminate the negative monomials of $\mathcal{F}$ by iterated linear changes of variables, and indeed Asy2 reports that this step (called ``preresolution'') fails for this integral.
Instead, we can first perform a blowup that decreases the degree of the polynomial defining the variety of $\mathcal{F}$, using the following change of variables,
\begin{eqnarray}
\label{eq:change_of_variables_three_loop_hidden}
    \alpha_0 = y_0\cdot \alpha_1,\qquad \alpha_2 = y_2\cdot \alpha_3,\qquad \alpha_4 = y_4\cdot \alpha_5,\qquad \alpha_6 = y_6\cdot \alpha_7,
\end{eqnarray}
the second Symanzik polynomial then becomes,
\begin{align}
\mathcal{F}(\x;\s) = (-s_{12}) (y_4-y_0)(y_6-y_2) \alpha_1\alpha_3\alpha_5\alpha_7 + (-s_{13})(y_2-y_0)(y_6-y_4) \alpha_1\alpha_3\alpha_5\alpha_7.
\end{align}

Next, we dissect the integral by imposing a strict hierarchy $y_i \ge y_j \ge y_k \ge y_l$ between the even-numbered parameters.
Considering all possible hierarchies will split the integral into $4!=24$ new integrals.
Finally, we re-map the boundaries of integration of each of the dissected integrals to $[0,\infty]$ by changing variables according to
\begin{align}
&y_i = z_i + z_j + z_k + z_l,&
&y_j = z_j + z_k + z_l,&
&y_k = z_k + z_l,&
&y_l = z_l.&
\end{align}
We obtain the following 24 polynomials,
\begin{align}
\label{eq:f_resolved}
\mathcal{F}_{1}(\boldsymbol{\alpha};\s) &= \alpha_1 \alpha_3 \alpha_5 \alpha_7 \left[ -s_{12}(\alpha_0 + \alpha_2)(\alpha_2+\alpha_4) - s_{13}(\alpha_0 \alpha_4)\right], \\
\mathcal{F}_{2}(\boldsymbol{\alpha};\s) &= \alpha_1 \alpha_3 \alpha_5 \alpha_7 \left[ -s_{12}(\alpha_2)(\alpha_0+\alpha_2+\alpha_6) + s_{13}(\alpha_0 \alpha_6)\right], \\
\mathcal{F}_{3}(\boldsymbol{\alpha};\s) &= \alpha_1 \alpha_3 \alpha_5 \alpha_7 \left[ -s_{12}(\alpha_0 \alpha_2) - s_{13}(\alpha_0 + \alpha_4) (\alpha_2 + \alpha_4)\right], \\
\mathcal{F}_{4}(\boldsymbol{\alpha};\s) &= \alpha_1 \alpha_3 \alpha_5 \alpha_7 \left[ s_{12}(\alpha_0 \alpha_6) - s_{13} (\alpha_4)(\alpha_0 + \alpha_4 + \alpha_6)\right], \\
\mathcal{F}_{5}(\boldsymbol{\alpha};\s) &= \alpha_1 \alpha_3 \alpha_5 \alpha_7 \left[ s_{12}(\alpha_6)(\alpha_0+\alpha_2+\alpha_6) + s_{13}(\alpha_0 + \alpha_6)(\alpha_2+\alpha_6)\right], \\
\mathcal{F}_{6}(\boldsymbol{\alpha};\s) &= \alpha_1 \alpha_3 \alpha_5 \alpha_7 \left[ s_{12}(\alpha_0 + \alpha_6)(\alpha_4+\alpha_6) + s_{13}(\alpha_6)(\alpha_0 + \alpha_4+ \alpha_6)\right],
\end{align}
with each of the remaining 18 integrals equal to one of the above up to a relabelling of the Feynman parameters.
The new $\mathcal{F}$ polynomials have only monomials of definite sign multiplying each of the invariants $s_{12}$ and $s_{13}$ and are therefore free from pinch singularities within the integration domain.

The dissected integrals can now be numerically evaluated using sector decomposition with a contour deformation.
Alternatively, it is possible to dramatically improve the numerical precision using the techniques described in Ref.~\cite{TomLoopsLegsProc}. 
Summing over the sectors, the full numeric result is given by,
\begin{align}
\mathcal{I}_{G_{\bullet \bullet}} &= 4\ (\I_1 + \I_2 + \I_3 + \I_4 + \I_5 + \I_6) \nonumber \\
&= \epsilon^{-4} \left[8.3400403920\mathbf{28} -  52.35987755983\mathbf{47} I \right] + \mathcal{O}(\epsilon^{-3}),
\end{align}
this agrees, within the numerical integration error, with the analytic result obtained in Ref.~\cite{Bargiela:2021wuy}.

Dissecting the integral, or the associated Newton polytope, has mapped the pinch surface which was originally within the domain of integration to the boundary of integration.
With the singularity now at the boundary, it can be resolved using the method of sector decomposition, enabling the numerical evaluation of the integral.

\section{Uncovering Hidden Regions}
\label{sec:hidden}

Landau singularities and the associated pinch surfaces are known to dictate the properties of asymptotic expansions of Feynman integrals.
In this section, we study the implication of pinch Landau singularities in parameter space from the perspective of the MoR.
We observe that they are associated with hidden regions which appear both in wide-angle scattering and the forward limit.

\subsection{On-shell expansion for wide-angle scattering}
\label{sec:onshell}

\begin{figure}[h]
\centering
\begin{subfigure}[b]{0.45\textwidth}
\centering
\resizebox{0.9\textwidth}{!}{
\begin{tikzpicture}[line width = 0.6, font=\large, mydot/.style={circle, fill, inner sep=.7pt}]
%\draw [help lines] (-1,1) grid (12,10);

\node[draw=Blue,circle,minimum size=1cm,fill=Blue!50] (h) at (6,8){};
\node[dashed, draw=Rhodamine,circle,minimum size=1.8cm,fill=Rhodamine!50] (s) at (3,2){};
\node[draw=Green,ellipse,minimum height=3cm, minimum width=1.1cm,fill=Green!50,rotate=-52] (j1) at (2,5){};
\node[draw=LimeGreen,ellipse,minimum height=3cm, minimum width=1.1cm,fill=LimeGreen!50] (j2) at (6,3.5){};
\node[draw=olive,ellipse,minimum height=3cm, minimum width=1.1cm,fill=olive!50,rotate=52] (jn) at (10,5){};

\node at (h) {$H$};
\node at (s) {\Large $S$};
\node at (j1) {\Large $J_1$};
\node at (j2) {\Large $J_2$};
\node at (jn) {\Large $J_K$};

\path (h) edge [double,double distance=2pt,color=Green] (j1) {};
\path (h) edge [double,double distance=2pt,color=LimeGreen] (j2) {};
\path (h) edge [double,double distance=2pt,color=olive] (jn) {};

\draw (s) edge [dashed,double,color=Rhodamine] (h) node [right] {};
\draw (s) edge [dashed,double,color=Rhodamine,bend left = 15] (j1) {};
\draw (s) edge [dashed,double,color=Rhodamine,bend right = 15] (j2) {};
\draw (s) edge [dashed,double,color=Rhodamine,bend right = 40] (jn) {};

\node (q1) at (4.5,9.5) {};
\node (q1p) at (4.7,9.7) {};
\node (qn) at (7.5,9.5) {};
\node (qnp) at (7.3,9.7) {};
\draw (q1) edge [color=Blue] (h) node [] {};
\draw (q1p) edge [color=Blue] (h) node [left] {$q_{K+1}$};
\draw (qn) edge [color=Blue] (h) node [] {};
\draw (qnp) edge [color=Blue] (h) node [right] {$q_M$};

\node (p1) at (0,3.5) {};
\node (p2) at (6,1) {};
\node (pn) at (12,3.5) {};
\draw (p1) edge [color=Green] (j1) node [below] {\Large $p_1$};
\draw (p2) edge [color=LimeGreen] (j2) node [below] {\Large $p_2$};
\draw (pn) edge [color=olive] (jn) node [below] {\Large $p_K$};

\path (j2)-- node[mydot, pos=.333] {} node[mydot] {} node[mydot, pos=.666] {}(jn);
\path (q1)-- node[mydot, pos=.333] {} node[mydot] {} node[mydot, pos=.666] {}(qn);

\end{tikzpicture}
}
\vspace{-2em}
\caption{Momentum configuration of a generic facet region}
\label{sfig:generic}
\end{subfigure}
\qquad
\begin{subfigure}[b]{0.45\textwidth}
\centering
\resizebox{0.65\textwidth}{!}{
\begin{tikzpicture}[line width = 0.6, font=\large, mydot/.style={circle, fill, inner sep=.7pt}, front->-/.style={decoration={markings,mark=at position 0.7 with {\arrow{Straight Barb[angle=45:2pt 8]}}}, postaction={decorate}}, half->-/.style={decoration={markings,mark=at position 0.5 with {\arrow{Straight Barb[angle=45:2pt 8]}}}, postaction={decorate}}, back->-/.style={decoration={markings,mark=at position 0.25 with {\arrow{Straight Barb[angle=45:2pt 8]}}}, postaction={decorate}}, transform shape]
% \draw [help lines] (0,0) grid (10,10);

\draw (0.5,1) edge [ultra thick, color=Green] (2,2) node [] {};
\draw (9.5,1) edge [ultra thick, color=teal] (8,2) node [] {};
\draw (0.5,9) edge [ultra thick, color=LimeGreen] (2,8) node [] {};
\draw (9.5,9) edge [ultra thick, color=olive] (8,8) node [] {};
\draw (5,6) edge [thick, double, double distance=2pt, color=Green, bend right = 10] (2,2) node [] {};
\draw (5,6) edge [thick, double, double distance=2pt, color=teal, bend left = 10] (8,2) node [] {};
\draw (5,6) edge [thick, double, double distance=2pt, color=LimeGreen, bend right = 10] (2,8) node [] {};
\draw (5,6) edge [thick, double, double distance=2pt, color=olive, bend left = 10] (8,8) node [] {};
\draw (2,2) edge [thick, draw=white, double=white, double distance=4pt, bend right = 10] (5,4) node [] {};\draw (2,2) edge [] [thick, double, double distance=2pt, color=Green, bend right = 10] (5,4) node [] {};
\draw (5,4) edge [thick, draw=white, double=white, double distance=4pt, bend right = 10] (8,2) node [] {};\draw (5,4) edge [] [thick, double, double distance=2pt, color=teal, bend right = 10] (8,2) node [] {};
\draw (2,8) edge [thick, draw=white, double=white, double distance=5pt, bend right = 10] (5,4) node [] {};\draw (2,8) edge [] [thick, double, double distance=2pt, color=LimeGreen, bend right = 10] (5,4) node [] {};
\draw (5,4) edge [thick, draw=white, double=white, double distance=5pt, bend right = 10] (8,8) node [] {};\draw (5,4) edge [] [thick, double, double distance=2pt, color=olive, bend right = 10] (8,8) node [] {};

\draw (5,9) edge [dashed, double, color=Rhodamine, bend left = 30] (7.5,7.5) node [] {};
\draw (5,9) edge [dashed, double, color=Rhodamine, bend right = 30] (2.5,7.5) node [] {};
\draw (5,9) edge [dashed, double, color=Rhodamine] (5,6) node [] {};

\node[draw=Green,ellipse,minimum height=3cm, minimum width=1.1cm,fill=Green!50,rotate=-50] (j1) at (2.5,2.5){};
\node[draw=teal,ellipse,minimum height=3cm, minimum width=1.1cm,fill=teal!50,rotate=50] (j1) at (7.5,2.5){};
\node[draw=LimeGreen,ellipse,minimum height=3cm, minimum width=1.1cm,fill=LimeGreen!50,rotate=50] (j1) at (2.5,7.5){};
\node[draw=olive,ellipse,minimum height=3cm, minimum width=1.1cm,fill=olive!50,rotate=-50] (j1) at (7.5,7.5){};
\node[draw=Rhodamine, dashed, circle, fill=Rhodamine!50, minimum size = 2cm] () at (5,9){};
\draw[fill, Blue!50, thick] (5,6) circle (15pt);\draw[Blue, thick] (5,6) circle (15pt);
\draw[fill, Blue!50, thick] (5,4) circle (15pt);\draw[Blue, thick] (5,4) circle (15pt);

\path (5,4.5)-- node[mydot, pos=.2] {} node[mydot] {} node[mydot, pos=.8] {}(5,5.5);
% \path (8,3)-- node[mydot, pos=.2] {} node[mydot] {} node[mydot, pos=.8] {}(8,7);

\node () at (0.5,0.5) {\huge $p_1$};
\node () at (9.5,0.5) {\huge $p_3$};
\node () at (0.5,9.5) {\huge $p_2$};
\node () at (9.5,9.5) {\huge $p_4$};
\node () at (2.5,2.5) {\huge $J_1$};
\node () at (7.5,2.5) {\huge $J_3$};
\node () at (2.5,7.5) {\huge $J_2$};
\node () at (7.5,7.5) {\huge $J_4$};
\node () at (5,4) {\Large $H_1$};
\node () at (5,6) {\Large $H_2$};
\node () at (5,9) {\huge $S$};
\end{tikzpicture}
}
\vspace{-1.5em}
\caption{Momentum configuration of the hidden region}
\label{sfig:hidden}
\end{subfigure}
\caption{The general configuration of facet and hidden regions in the on-shell expansion of $2\to 2$ wide-angle scattering. 
The graph $G$ is the union of the hard subgraph $H$, the jet subgraphs $J_1,\dots, J_4$, and the soft subgraph $S$. 
For the hidden region, in contrast to the facet regions, the hard subgraph has multiple connected components $H_1, H_2,\dots$, and all four jets are attached to each of these components.}
\label{fig:wide_angle_scattering_hidden_region}
\end{figure}
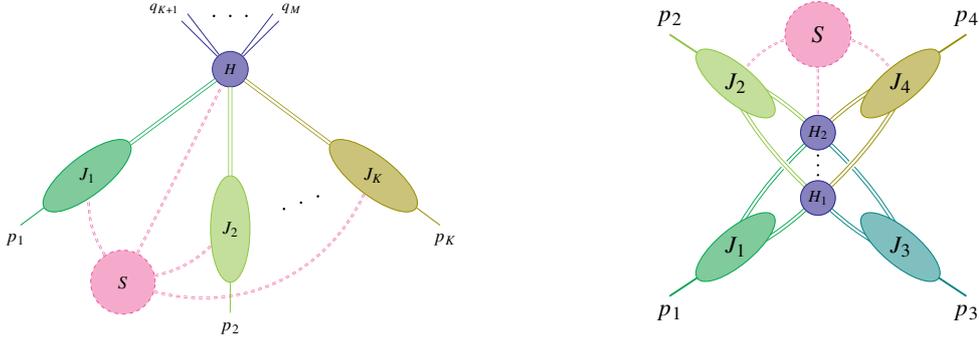

Let us examine the presence of hidden regions in the on-shell expansion of wide-angle scattering.
This limit is defined by considering an off-shell Green's function with $M$ external legs, out of which $K_0$ are strictly on-shell, while $K_\lambda$ are expanded about the on-shell limit, i.e.
\begin{eqnarray}
\label{eq:wideangle_onshell_kinematics}
\left\lbrace 
\begin{array}{ll}
p_i^2=0 
& \qquad i=1,\dots,K_0
\\
p_i^2\sim \lambda Q^2 &\qquad i=K_0+1,\dots,K,
\\
q_i^2\sim Q^2 
&\qquad i=K+1,\dots,M
\\
p_{i_1}\cdot p_{i_2}\sim Q^2\ \ 
&\qquad \forall \,\, i_1\neq i_2\,,
\end{array}
\right.
\end{eqnarray}
with $\lambda \rightarrow 0$.
Here, we consider only diagrams containing massless propagators.
The wide-angle condition is incorporated in $p_{i_1}\cdot p_{i_2}\sim Q^2$, implying that the angle between the three-momenta $\boldsymbol{p}_{i_1}$ and $\boldsymbol{p}_{i_2}$ is $\mathcal{O}(1)$ for any $i_{1}$ and~$i_{2}$.
Given any massless scattering graph with the kinematics of~(\ref{eq:wideangle_onshell_kinematics}), the complete list of facet regions can be described by the hard-collinear-soft picture in Figure~\ref{sfig:generic}, as proposed in Ref.~\cite{GrdHzgJnsMaSchlk22} and proved in Ref.~\cite{Ma23}.

Following the work of Ref.~\cite{BottsStm89}, let us consider graph $G_{\bullet \bullet}$ in the limit,
\begin{align}
p_i^2 = \lambda Q^2,\qquad 
p_i\cdot v_i\sim \lambda Q, \qquad 
p_i\cdot \overline{v}_i\sim Q,\qquad 
p_i\cdot v_{i\perp} \sim \sqrt{\lambda} Q.
\end{align}
Introducing a fourth, fully constrained, loop momentum, $k_4$, and considering the mode with all $k_i$ collinear to $p_i$, we can parameterise the loop momenta as,
\begin{align}
\label{eq:parameterisation_jet_like_k_i}
    k_i^\mu = Q\left(
    \xi_i v_i^\mu + \lambda \kappa_i\overline{v}_i^\mu 
    + \sqrt{\lambda}\tau_i u_i^\mu 
    + \sqrt{\lambda}\nu_i n^\mu_i \right),\qquad i=1,2,3,4.
\end{align}
Changing integration variables from the individual components of $k_i$ to $\prod_{i=1}^3 \left(\mathrm{d} \kappa_i \mathrm{d} \tau_i \mathrm{d} \nu_i \right) \mathrm{d} \kappa_4 \mathrm{d} \tau_4$, the scaling of the integration measure is $\lambda^{6-3\epsilon} \lambda^{3/2}$.
This suggests the presence of a leading region (the Landshoff scattering region) with degree of divergence,
\begin{align}
    \mu[G_{\bullet\bullet},\text{Landshoff}] = 6-3\epsilon +\frac{3}{2} -8 = -\frac{1}{2}-3\epsilon.
\end{align}

\begin{table}[h]
    \small
    \centering
    \begin{tabular}{l|l}
        \bf $\v_\mathrm{R}$ ($x_0, x_1, \ldots, x_7 ;\lambda$)      & \bf degree of divergence \\\hline
        $(-2,-1,-2,-1,-2,-1,-2,-1;1)$ & $-6\epsilon$\\
        $(-1,-2,-1,-2,-1,-2,-1,-2;1)$ & $-6\epsilon$\\
        $(-1,-1,-1,0,-1,0,-1,0;1)$    & $1-3\epsilon$\\
        $(-1,-1,0,-1,0,-1,0,-1;1)$    & $1-3\epsilon$\\
        $(-1,-1,0,0,0,0,0,0;1)$       & $-\epsilon$\\
        $(0,0,0,0,0,0,0,0;1)$         & $0$\\
    \end{tabular}
    \caption{Regions obtained by directly applying the MoR to the on-shell expansion of graph $G_{\bullet \bullet}$ with $p_1^2 \sim \lambda Q^2$ and $p_i^2=0\ (i=2,3,4)$.}
    \label{tab:regions_onshell_original}
\end{table} 

We can now check for the presence of the Landshoff region in parameter space.
Using the geometric formulation of the MoR we can directly derive the facet regions from the Synamzik polynomials of $G_{\bullet \bullet}$ given Eq.~\eqref{eq:UF_Gbb}.
For brevity, we report in Table~\ref{tab:regions_onshell_original} the regions obtained when considering the expansion $p_i^2 \sim \lambda Q^2$ and $p_i^2=0 (i=2,3,4)$.
We note that the Landshoff region is not detected as a facet region of the original Newton polytope.
According to this analysis, the leading power behaviour of the facet regions of the integral in the small $\lambda$ limit is $\mathcal{O}(\lambda^0)$.

\begin{table}[h]
    \small
    \centering
    \begin{tabular}{l|l|l}
        \multirow{2}{5.5cm}{\bf $\v_\mathrm{R}$ ($y_0, x_1, y_2, x_3, y_4, x_5, y_6, x_7 ;\lambda$)} &  \multirow{2}{5cm}{\bf $\v_\mathrm{R}$ ($x_0, x_1, \ldots, x_7; \lambda$)} & \bf degree of \\
         & & \bf divergence \\\hline
        $(1/2,-1,1/2,-1,1/2,-1,0,-1;1)$       & $(-2,-2,-2,-2,-2,-2,-2,-2;2)$ & $-1/2-3\epsilon$\\
        $(0,-1,1,-1,1,-1,0,-1;1)$    & $(-1,-1,-1,-1,-1,-1,-1,-1;1)$ & $-3\epsilon$\\
        $(1,-1,1,-1,0,-1,0,-1;1)$         & $(-1,-1,-1,-1,-1,-1,-1,-1;1)$ & $-3\epsilon$\\
        $(-1,-1,-1,-1,-1,-1,-1,-1;1)$ & $(-2,-1,-2,-1,-2,-1,-2,-1;1)$ & $-6\epsilon$\\
        $(1,-2,1,-2,1,-2,1,-2;1)$         & $(-1,-2,-1,-2,-1,-2,-1,-2;1)$ & $-6\epsilon$\\
        $(0,-1,0,0,0,0,0,0;1)$ & $(-1,-1,0,0,0,0,0,0;1)$ & $-\epsilon$\\
        $(0,0,0,0,0,0,0,0;1)$    & $(0,0,0,0,0,0,0,0;1)$ & $0$\\
    \end{tabular}
    \caption{On-shell expansion of $G_{\bullet \bullet}$, with $p_1^2 \sim \lambda Q^2$ and $p_i^2=0\ (i=2,3,4)$, first integral dissection $\mathcal{I}_1$.}
    \label{tab:regions_onshell_split1}
\end{table}

Alternatively, we can dissect the polytope using the procedure described in Section.~\ref{sec:evaluating}. 
We obtain a set of 24 new integrals, $\mathcal{I}_1, \ldots, \mathcal{I}_{24}$.
These integrals have $\mathcal{F}$ polynomials similar to Eq.~\eqref{eq:f_resolved}, but with additional terms proportional to $(-p_1^2)$ multiplied by non-negative polynomials of the parameters.
Each of the new integrals has a same-sign regime for some choice of the value of $p_1^2, s$ and $t$ (although not the same choice for all integrals). 
We, therefore, expect that each integral separately can be analytically continued from a same-sign regime to the region of interest, and the application of the MoR should now identify all scaleful regions, i.e. there are no hidden regions in the new integrals.
In Table~\ref{tab:regions_onshell_split1} we show the regions obtained by analysing the first integral dissection $\mathcal{I}_1$.
We observe that the Landshoff region is present in all 24 integral dissections and that it is the leading-power region in the small-$\lambda$ limit, with the degree of divergence $\mu = -1/2 -3 \epsilon$, as expected from the momentum space analysis.
In Ref.~\cite{Gardi:2024axt}, we show by direct numerical evaluation that the Landshoff region reproduces the correct behaviour of the full integral in the small $\lambda$ limit.

In Figure~\ref{sfig:generic} we show the possible momentum configurations obeyed by generic facet regions of integrals in the on-shell expansion, as described in Refs.~\cite{GrdHzgJnsMaSchlk22,Ma23}, and in Figure~\ref{sfig:hidden} we show the momentum configuration of the Landshoff region.
We observe that, unlike generic facet regions, the hard subgraph of the hidden region has multiple connected components.

\subsection{Regge-limit expansion for $2 \rightarrow 2$ scattering }
\label{sec:regge}

Let us now turn our attention to the Regge-limit expansion for $2\to 2$ scattering, where every external momentum is strictly massless and on shell, with $p_3$ (nearly) collinear to $p_1$ and $p_4$ nearly collinear to $p_2$, namely,
\begin{eqnarray}
\label{Regge_limit}
    p_i^2=0\ \ (i=1,2,3,4),\qquad  \frac{-t}{s} \sim  \frac{-t}{-u} \sim \lambda,
\end{eqnarray}
where $s\equiv s_{12}\equiv (p_1+p_2)^2>0$, $t\equiv s_{13}\equiv  (p_1-p_3)^2<0$, and $u\equiv s_{14} \equiv (p_1-p_4)^2<0$. 
In contrast to the wide-angle scattering kinematics, eq.~(\ref{eq:wideangle_onshell_kinematics}), the Regge-limit expansion is known to feature Glauber modes that contribute to the region structure in certain graphs.

We may repeat the momentum space analysis presented in Section~\ref{sec:onshell} in the case of forward scattering.
However, for the three-loop integrals considered here, complete analytic results are known when all external legs are exactly on shell~\cite{Henn:2020lye,Bargiela:2021wuy}.
We therefore have the opportunity to directly verify the results we obtain via asymptotic expansions against the known analytic results.

\begin{table}[h]
    \small
    \centering
    \begin{tabular}{l|l}
        \bf $\v_\mathrm{R}$ ($x_0, x_1, \ldots, x_7 ;\lambda$)      & \bf degree of divergence \\\hline
        $(-1,-1,-1,0,-1,-1,-1,0;1)$ & $-3\epsilon$\\
        $(-1,-1,0,-1,-1,-1,0,-1;1)$ & $-3\epsilon$\\
        $(-1,0,-1,-1,-1,0,-1,-1;1)$    & $-3\epsilon$\\
        $(0,-1,-1,-1,0,-1,-1,-1;1)$    & $-3\epsilon$\\
        $(0,0,0,0,0,0,0,0;1)$         & $0$\\
    \end{tabular}
    \caption{Regions obtained by directly applying the MoR to the Regge-limit expansion of graph $G_{\bullet \bullet}$.}
    \label{tab:regions_forward_original}
\end{table}

In the notation of ref.~\cite{Bargiela:2021wuy}, the graph $G_{\bullet\bullet}$ is written as
\[
J_{G_{\bullet\bullet}}(s_{12},s_{13};\epsilon)=\tt{{\rm INT}[``{\rm NPL2}", 8, 4009, 8, 0, {1, 0, 0, 1, 0, 1, 0, 1, 1, 1, 1, 1, 0, 0, 0}]}.
\]
Expanding the analytic result about the forward limit, with $x = -s_{13}/s_{12}$, we observe that
\begin{align}
J_{G_{\bullet\bullet}}(x;\epsilon)\sim x^{-1-3\epsilon}\quad\qquad \text{for}\quad x\to 0\,.
\end{align}
In contrast, directly applying the MoR to $G_{\bullet \bullet}$ in the forward limit, we obtain the set of region vectors shown in Table~\ref{tab:regions_forward_original}.
We observe that the leading-power region behaves only as $\mathcal{O}(\lambda^0)$ and the leading region is missed.

\begin{table}[h]
    \small
    \centering
    \begin{tabular}{l|l|l}
        \multirow{2}{5.5cm}{\bf $\v_\mathrm{R}$ ($y_0, x_1, y_2, x_3, y_4, x_5, y_6, x_7 ;\lambda$)} &  \multirow{2}{5cm}{\bf $\v_\mathrm{R}$ ($x_0, x_1, \ldots, x_7; \lambda$)} & \bf degree of \\
         & & \bf divergence \\\hline
        $(0,-1,0,-1,0,-1,1,-1;1)$    & $(-1,-1,-1,-1,-1,-1,-1,-1;1)$ & $-1-3\epsilon$\\
        $(1,-1,0,-1,0,-1,0,-1;1)$ & $(-1,-1,-1,-1,-1,-1,-1,-1;1)$ & $-1-3\epsilon$\\
        $(-1,0,0,-1,-1,0,0,-1;1)$       & $(-1,0,-1,-1,-1,0,-1,-1;1)$ & $-3\epsilon$\\
        $(0,0,0,0,0,0,0,0;1)$         & $(0,0,0,0,0,0,0,0;1)$ & $0$\\
    \end{tabular}
    \caption{Regge-limit expansion of graph $G_{\bullet \bullet}$, fourth integral dissection $\mathcal{I}_4$. The region $(-1,-1,-1,0,-1,-1,-1,0;1)$ in the original variables, is present in other integral dissections, e.g. $\mathcal{I}_{11}$. The region $(0,-1,-1,-1,0,-1,-1,-1;1)$ in the original polytope, is entirely absent after dissection.}
    \label{tab:regions_forward_split1} 
\end{table}

Following the resolution procedure described in Section~\ref{sec:evaluating}, we again obtain a set of 24 new integrals.
By the same argument as presented in Section~\ref{sec:onshell}, we reason that the new integrals should not contain hidden regions.
In Table~\ref{tab:regions_forward_split1}, we present the regions obtained for the integral dissection $\mathcal{I}_4$, we observe that indeed regions with degree of divergence $\mu=-1-3\epsilon$ are present, as expected from the expansion of the analytic result.
Therefore, by dissecting the Newton polytope of the original integral, we have exposed the hidden region and recovered the correct expansion of the integral in the forward region.
Using the momentum space analysis, we can show that the interpretation of the hidden region is compatible with the exchanged loop momenta obeying the Glauber mode scaling law $k\sim Q(\lambda,\lambda,\lambda^{1/2})$.
We emphasise, however, that in general, the Glauber mode does not feature only in hidden regions and can also be present in facet regions.
The rich mode and region structure of the Regge limit at three loops and beyond will be investigated in more detail in a forthcoming publication~\cite{GrdHzgJnsMaprepare}.

\section{Conclusion \& Outlook}
\label{sec:conclusions}

Recognising the value of parametric representations of Feynman integrals and their geometrical interpretation in treating singularities as endpoint divergences, here, we focused on the exception: singularities which manifest themselves as pinches in parameter space, and obstruct the application of existing strategies based on the Newton polytope. 
Earlier work in this direction~\cite{JtzSmnSmn12,AnthnrySkrRmn19}, in the context of the MoR, focused on linear cancellation between pairs of Feynman parameters, while our analysis identified more general cancellation patterns, including higher-degree polynomials involving multiple Feynman parameters.
We observed that such cancellations can occur not just in special kinematic limits such as the threshold expansion and forward limit, but also in more general situations where the Mandelstam invariants are less constrained (e.g. wide-angle scattering).

We have presented an algorithm which aids in identifying graphs in which a pinch in parameter space can occur.
Such pinches can hinder the Newton-polytope based sector decomposition algorithms and obstruct the determination of a complete set of regions in the MoR.
Next, we addressed the numerical computation of integrals which feature a pinch in parameter space within the domain of integration. 
Although the straightforward application of existing sector decomposition tools is doomed to fail due to the presence of such pinch singularities, we showed how a procedure, based on dissecting the Newton polytope, allows their evaluation.
Finally, we addressed the determination of the complete set of regions needed for asymptotic expansions of integrals around a limit in which a pinch in parameter space occurs.

\section*{Acknowledgments}

We would like to thank the organisers of the ``Loops and Legs in Quantum Field Theory'' conference for the opportunity to present this work.
We also thank Charalampos Anastasiou, Thomas Becher, Holmfridur Hannesdottir, Andrew McLeod, Erik Panzer, Johannes Schlenk, and George Sterman for useful discussions and Piotr Bargiela for his valuable help with the analytic three-loop four-point results.
EG and FH are supported by the STFC Consolidated Grant ``Particle Physics at the Higgs Centre''. FH is also supported by the UKRI FLF grant ``Forest Formulas for the LHC'' (Mr/S03479x/1). 
SJ is supported by the Royal Society University Research Fellowship (URF/R1/201268) and by the UK Science and Technology Facilities Council under contract ST/T001011/1.
YM is supported by the Swiss National Science Foundation through its project funding scheme, grant number 10001706.

\bibliographystyle{JHEP}
\bibliography{main}

\providecommand{\href}[2]{#2}\begingroup\raggedright\begin{thebibliography}{10}

\bibitem{Lnd59}
L.~Landau, {\it On analytic properties of vertex parts in quantum field
  theory},  {\em Nuclear Physics} {\bf 13} (1959), no.~1 181--192.

\bibitem{Bjorken:1959fd}
J.~D. Bjorken, {\em {Experimental tests of Quantum electrodynamics and spectral
  representations of Green's functions in perturbation theory}}.
\newblock PhD thesis, Stanford U., 1959.

\bibitem{Nakanishi}
N.~Nakanishi, {\it {Ordinary and Anomalous Thresholds in Perturbation Theory}},
   {\em Progress of Theoretical Physics} {\bf 22} (07, 1959) 128--144,
  [\href{http://xxx.lanl.gov/abs/https://academic.oup.com/ptp/article-pdf/22/1/128/5427385/22-1-128.pdf}{{\tt
  https://academic.oup.com/ptp/article-pdf/22/1/128/5427385/22-1-128.pdf}}].

\bibitem{EdenLdshfOlvPkhn02book}
R.~J. Eden, P.~V. Landshoff, D.~I. Olive, and J.~C. Polkinghorne, {\em The
  analytic S-matrix}.
\newblock Cambridge University Press, 2002.

\bibitem{Brown:2009ta}
F.~C.~S. Brown, {\it {On the periods of some Feynman integrals}},
  \href{http://xxx.lanl.gov/abs/0910.0114}{{\tt arXiv:0910.0114}}.

\bibitem{Panzer:2014caa}
E.~Panzer, {\it {Algorithms for the symbolic integration of hyperlogarithms
  with applications to Feynman integrals}},  {\em Comput. Phys. Commun.} {\bf
  188} (2015) 148--166, [\href{http://xxx.lanl.gov/abs/1403.3385}{{\tt
  arXiv:1403.3385}}].

\bibitem{GrdHzgJnsMaSchlk22}
E.~Gardi, F.~Herzog, S.~Jones, Y.~Ma, and J.~Schlenk, {\it {The on-shell
  expansion: from Landau equations to the Newton polytope}},  {\em JHEP} {\bf
  07} (2023) 197, [\href{http://xxx.lanl.gov/abs/2211.14845}{{\tt
  arXiv:2211.14845}}].

\bibitem{FvlMzrTln23pld}
C.~Fevola, S.~Mizera, and S.~Telen, {\it {Principal Landau determinants}},
  {\em Comput. Phys. Commun.} {\bf 303} (2024) 109278,
  [\href{http://xxx.lanl.gov/abs/2311.16219}{{\tt arXiv:2311.16219}}].

\bibitem{FvlMzrTln23prl}
C.~Fevola, S.~Mizera, and S.~Telen, {\it {Landau Singularities Revisited:
  Computational Algebraic Geometry for Feynman Integrals}},  {\em Phys. Rev.
  Lett.} {\bf 132} (2024), no.~10 101601,
  [\href{http://xxx.lanl.gov/abs/2311.14669}{{\tt arXiv:2311.14669}}].

\bibitem{Dlapa:2023cvx}
C.~Dlapa, M.~Helmer, G.~Papathanasiou, and F.~Tellander, {\it {Symbol alphabets
  from the Landau singular locus}},  {\em JHEP} {\bf 10} (2023) 161,
  [\href{http://xxx.lanl.gov/abs/2304.02629}{{\tt arXiv:2304.02629}}].

\bibitem{Helmer:2024wax}
M.~Helmer, G.~Papathanasiou, and F.~Tellander, {\it {Landau Singularities from
  Whitney Stratifications}},  \href{http://xxx.lanl.gov/abs/2402.14787}{{\tt
  arXiv:2402.14787}}.

\bibitem{LeePmrsk13}
R.~N. Lee and A.~A. Pomeransky, {\it {Critical points and number of master
  integrals}},  {\em JHEP} {\bf 11} (2013) 165,
  [\href{http://xxx.lanl.gov/abs/1308.6676}{{\tt arXiv:1308.6676}}].

\bibitem{AkHmHlmMzr22}
N.~Arkani-Hamed, A.~Hillman, and S.~Mizera, {\it {Feynman polytopes and the
  tropical geometry of UV and IR divergences}},  {\em Phys. Rev. D} {\bf 105}
  (2022), no.~12 125013, [\href{http://xxx.lanl.gov/abs/2202.12296}{{\tt
  arXiv:2202.12296}}].

\bibitem{SmnTtyk09FIESTA}
A.~Smirnov and M.~Tentyukov, {\it Feynman integral evaluation by a sector
  decomposition approach (fiesta)},  {\em Computer Physics Communications} {\bf
  180} (2009), no.~5 735--746.

\bibitem{SmnSmnTtyk11FIESTA2}
A.~Smirnov, V.~Smirnov, and M.~Tentyukov, {\it Fiesta 2: parallelizeable
  multiloop numerical calculations},  {\em Computer Physics Communications}
  {\bf 182} (2011), no.~3 790--803.

\bibitem{Smn14FIESTA3}
A.~V. Smirnov, {\it Fiesta 3: cluster-parallelizable multiloop numerical
  calculations in physical regions},  {\em Computer Physics Communications}
  {\bf 185} (2014), no.~7 2090--2100.

\bibitem{Smn16FIESTA4}
A.~V. Smirnov, {\it Fiesta 4: Optimized feynman integral calculations with gpu
  support},  {\em Computer Physics Communications} {\bf 204} (2016) 189--199.

\bibitem{Smn22FIESTA5}
A.~Smirnov, N.~Shapurov, and L.~Vysotsky, {\it Fiesta5: numerical
  high-performance feynman integral evaluation},  {\em Computer Physics
  Communications} {\bf 277} (2022) 108386.

\bibitem{KnkUeda10}
T.~Kaneko and T.~Ueda, {\it A geometric method of sector decomposition},  {\em
  Computer Physics Communications} {\bf 181} (2010), no.~8 1352--1361.

\bibitem{Carter:2010hi}
J.~Carter and G.~Heinrich, {\it {SecDec: A general program for sector
  decomposition}},  {\em Comput. Phys. Commun.} {\bf 182} (2011) 1566--1581,
  [\href{http://xxx.lanl.gov/abs/1011.5493}{{\tt arXiv:1011.5493}}].

\bibitem{Borowka:2015mxa}
S.~Borowka, G.~Heinrich, S.~P. Jones, M.~Kerner, J.~Schlenk, and T.~Zirke, {\it
  {SecDec-3.0: numerical evaluation of multi-scale integrals beyond one loop}},
   {\em Comput. Phys. Commun.} {\bf 196} (2015) 470--491,
  [\href{http://xxx.lanl.gov/abs/1502.06595}{{\tt arXiv:1502.06595}}].

\bibitem{Heinrich:2023til}
G.~Heinrich, S.~P. Jones, M.~Kerner, V.~Magerya, A.~Olsson, and J.~Schlenk,
  {\it {Numerical scattering amplitudes with pySecDec}},  {\em Comput. Phys.
  Commun.} {\bf 295} (2024) 108956,
  [\href{http://xxx.lanl.gov/abs/2305.19768}{{\tt arXiv:2305.19768}}].

\bibitem{pySecDec17}
S.~Borowka, G.~Heinrich, S.~Jahn, S.~P. Jones, M.~Kerner, J.~Schlenk, and
  T.~Zirke, {\it {pySecDec: a toolbox for the numerical evaluation of
  multi-scale integrals}},  {\em Comput. Phys. Commun.} {\bf 222} (2018)
  313--326, [\href{http://xxx.lanl.gov/abs/1703.09692}{{\tt
  arXiv:1703.09692}}].

\bibitem{Brsk20}
M.~Borinsky, {\it {Tropical Monte Carlo quadrature for Feynman integrals}},
  {\em Ann. Inst. H. Poincare D Comb. Phys. Interact.} {\bf 10} (2023), no.~4
  635, [\href{http://xxx.lanl.gov/abs/2008.12310}{{\tt arXiv:2008.12310}}].

\bibitem{BrskMchTld23}
M.~Borinsky, H.~J. Munch, and F.~Tellander, {\it {Tropical Feynman integration
  in the Minkowski regime}},  {\em Comput. Phys. Commun.} {\bf 292} (2023)
  108874, [\href{http://xxx.lanl.gov/abs/2302.08955}{{\tt arXiv:2302.08955}}].

\bibitem{LiuMa23AMFlow}
X.~Liu and Y.-Q. Ma, {\it {AMFlow: A Mathematica package for Feynman integrals
  computation via auxiliary mass flow}},  {\em Comput. Phys. Commun.} {\bf 283}
  (2023) 108565, [\href{http://xxx.lanl.gov/abs/2201.11669}{{\tt
  arXiv:2201.11669}}].

\bibitem{Hdg21DiffExp}
M.~Hidding, {\it {DiffExp, a Mathematica package for computing Feynman
  integrals in terms of one-dimensional series expansions}},  {\em Comput.
  Phys. Commun.} {\bf 269} (2021) 108125,
  [\href{http://xxx.lanl.gov/abs/2006.05510}{{\tt arXiv:2006.05510}}].

\bibitem{AmdlBcnDvtRanaVcn23SeaSyde}
T.~Armadillo, R.~Bonciani, S.~Devoto, N.~Rana, and A.~Vicini, {\it {Evaluation
  of Feynman integrals with arbitrary complex masses via series expansions}},
  {\em Comput. Phys. Commun.} {\bf 282} (2023) 108545,
  [\href{http://xxx.lanl.gov/abs/2205.03345}{{\tt arXiv:2205.03345}}].

\bibitem{JtzSmnSmn12}
B.~Jantzen, A.~V. Smirnov, and V.~A. Smirnov, {\it Expansion by regions:
  revealing potential and glauber regions automatically},  {\em The European
  Physical Journal C} {\bf 72} (2012), no.~9 1--14.

\bibitem{AnthnrySkrRmn19}
B.~Ananthanarayan, A.~Pal, S.~Ramanan, and R.~Sarkar, {\it Unveiling regions in
  multi-scale feynman integrals using singularities and power geometry},  {\em
  The European Physical Journal C} {\bf 79} (2019), no.~1 1--20.

\bibitem{HrchJnsSlk22}
G.~Heinrich, S.~Jahn, S.~Jones, M.~Kerner, F.~Langer, V.~Magerya, A.~Poldaru,
  J.~Schlenk, and E.~Villa, {\it Expansion by regions with pysecdec},  {\em
  Computer Physics Communications} {\bf 273} (2022) 108267.

\bibitem{Gardi:2024axt}
E.~Gardi, F.~Herzog, S.~Jones, and Y.~Ma, {\it {Dissecting polytopes: Landau
  singularities and asymptotic expansions in $2\to 2$ scattering}},
  \href{http://xxx.lanl.gov/abs/2407.13738}{{\tt arXiv:2407.13738}}.

\bibitem{Mdst63}
S.~Mandelstam, {\it {Cuts in the Angular Momentum Plane. 2}},  {\em Nuovo Cim.}
  {\bf 30} (1963) 1148--1162.

\bibitem{Hld64}
I.~G. Halliday, {\it {High-energy behavior at fixed angle in perturbation
  theory}},  {\em Annals Phys.} {\bf 28} (1964) 320--345.

\bibitem{Ldsf74}
P.~V. Landshoff, {\it {Model for elastic scattering at wide angle}},  {\em
  Phys. Rev. D} {\bf 10} (1974) 1024--1030.

\bibitem{BottsStm89}
J.~Botts and G.~F. Sterman, {\it {Hard Elastic Scattering in QCD: Leading
  Behavior}},  {\em Nucl. Phys. B} {\bf 325} (1989) 62--100.

\bibitem{TomLoopsLegsProc}
S.~Jones, A.~Olsson, and T.~Stone, {\it {Evaluating Parametric Integrals in the
  Minkowski Regime without Contour Deformation}},  in {\em {Loops and Legs in
  Quantum Field Theory}}, 7, 2024.
\newblock \href{http://xxx.lanl.gov/abs/2407.06973}{{\tt arXiv:2407.06973}}.

\bibitem{Bargiela:2021wuy}
P.~Bargiela, F.~Caola, A.~von Manteuffel, and L.~Tancredi, {\it {Three-loop
  helicity amplitudes for diphoton production in gluon fusion}},  {\em JHEP}
  {\bf 02} (2022) 153, [\href{http://xxx.lanl.gov/abs/2111.13595}{{\tt
  arXiv:2111.13595}}].

\bibitem{Ma23}
Y.~Ma, {\it {Identifying regions in wide-angle scattering via graph-theoretical
  approaches}},  \href{http://xxx.lanl.gov/abs/2312.14012}{{\tt
  arXiv:2312.14012}}.

\bibitem{Henn:2020lye}
J.~Henn, B.~Mistlberger, V.~A. Smirnov, and P.~Wasser, {\it {Constructing d-log
  integrands and computing master integrals for three-loop four-particle
  scattering}},  {\em JHEP} {\bf 04} (2020) 167,
  [\href{http://xxx.lanl.gov/abs/2002.09492}{{\tt arXiv:2002.09492}}].

\bibitem{GrdHzgJnsMaprepare}
E.~Gardi, F.~Herzog, S.~Jones, and Y.~Ma, {\it {Regions in the Regge limit of
  two to two scattering}},  {\em -- in preparation}.

\end{thebibliography}\endgroup

\end{document}